\documentclass[prb,showpacs,superscriptaddress,twocolumn,preprintnumbers,amssymb,amsfonts]{revtex4}

\pdfoutput=1

\usepackage{graphicx}
\usepackage{amssymb}
\usepackage{amsmath}
\usepackage{bm}
\usepackage{feynmf}
\usepackage{latexsym}
\usepackage{amsmath}
\usepackage{amssymb}
\usepackage{braket}
\usepackage{array}
\usepackage{ifthen}
\usepackage{verbatim}
\usepackage[english]{babel}
\usepackage{enumerate}
\usepackage{accents}
\usepackage{nicefrac}
\usepackage[utf8x]{inputenc}

\usepackage[T1]{fontenc}
\usepackage{cancel}
\usepackage{amsfonts}
\usepackage{mathrsfs}
\usepackage{braket}
\usepackage{mathdots}

\newcommand{\tr}[2]{\mathrm{Tr}_{#1}\left[#2\right]}
\newcommand{\be}{\begin{equation}}
\newcommand{\ee}{\end{equation}}

\begin{document}

\input epsf
\def\beq{\begin{equation}}
\def\eeq{\end{equation}}
\def\beqn{\begin{eqnarray}}
\def\eeqn{\end{eqnarray}}
\def\etal{\emph{et al.}}
\renewcommand{\v}[1]{\boldsymbol{#1}}
\newcommand{\up}{\uparrow}
\newcommand{\down}{\downarrow} 

\title{Entanglement spectrum of random-singlet quantum critical points}
\author{Maurizio Fagotti$^1$, Pasquale  Calabrese}
\affiliation{Dipartimento di Fisica dell'Universit\`a di Pisa and INFN, Pisa, Italy.}
\author{Joel~E.~Moore}
\affiliation{Department of Physics, University of California,
Berkeley, CA 94720} \affiliation{Materials Sciences Division,
Lawrence Berkeley National Laboratory, Berkeley, CA 94720}

\date{\today}

\begin{abstract}
The entanglement spectrum, i.e., the full distribution of Schmidt eigenvalues of the reduced density matrix, contains more information than the conventional entanglement entropy and has been studied recently in several many-particle systems.  We compute the disorder-averaged entanglement spectrum, in the form of the disorder-averaged moments $\overline{\rm{Tr} \rho_A^\alpha}$ of the reduced density matrix $\rho_A$, for a contiguous block of many spins at the random-singlet quantum critical point in one dimension.  The result compares well in the scaling limit with numerical studies on the random XX model and is also expected to describe the (interacting) random Heisenberg model.  Our numerical studies on the XX case reveal that the dependence of the entanglement entropy and spectrum on the geometry of the Hilbert space partition is quite different than for conformally invariant critical points.
\end{abstract}
\maketitle

\section{Introduction}

The ground state of a system at a quantum critical point shows universal behavior in many quantities.  Correlation functions, 
for example, show universal power-law behavior, and in some cases these power laws can be obtained exactly by mapping the 
quantum critical point to a system in one more dimension.  The most powerful example of this mapping is for one-dimensional (1D) quantum critical points (QCPs) that become 2D classical critical points with conformal invariance.  In addition to standard correlation functions, it is now understood that the entanglement entropy, reviewed below (see the comprehensive reviews in Ref. \onlinecite{rev}), 
is universal at such quantum critical points, and determined by the central charge of the associated 2D conformal field theory~\cite{holzhey,vidalent,cc-04,cc-rev}.  
For a partition of an infinite 1D system into a finite chain of length $\ell$ and the remainder, the entanglement entropy 
(the von Neumann entropy of the reduced density matrix $\rho_A$) for 
$\ell$ much larger than the short-distance cutoff $a$ is asymptotically
\be
S_{VN} \equiv-\tr{}{\rho_A\ln\rho_A}=\frac{c}3 \ln \frac{\ell}a +c'_1\,,
\label{criticalent}
\ee
where $c$ is the central charge and $c'_1$ a non-universal additive constant.

Other properties related to entanglement are less well understood, even at these quantum critical points, such as the entanglement spectrum (the full set of reduced density matrix eigenvalues) and the full set of entanglement Renyi entropies; one exception is free Fermi models, where the entanglement spectrum is given by the spectrum of an effective ``entanglement Hamiltonian''~\cite{peschelfermion}.  A form for the spectrum~\cite{calabreselefevre} at 1D conformal QCPs that is exact in some cases and a good approximation in others~\cite{pollmannmoore,franchini} can be used to develop a theory of how finite entanglement perturbs criticality in numerical studies~\cite{tagliacozzo,pollmannentanglement}.  
The entanglement spectrum has also been applied to understanding gapped (non-critical) topological 
phases~\cite{lihaldane,bernevig,lauchli}, where it contains information about the edge excitation spectrum that goes beyond 
the universal constant in the entanglement entropy~\cite{kitaevpreskill,levinwen,schoutens}. 
The same is true for quantum 2D models with conformal invariant ground-state wave-functions \cite{2d}.
Also results for a critical non-conformal 1D model are available \cite{ps-10}.

This paper studies the entanglement spectrum at ``random-singlet''  1D QCPs, in which quenched disorder leads to an RG flow to infinite randomness.   We obtain the disorder-averaged moments of the Schmidt eigenvalue distribution analytically and compare them to numerical results on a special case with a free-fermion representation, the random XX model.  While these critical points are not conformally invariant (after mapping to a 2D problem, the imaginary-time direction has no randomness and is hence very different from the spatial direction), their disorder-averaged correlation functions have nevertheless been understood in many cases~\cite{fisher1,fisher2,damledynamics} by real-space renormalization group method~\cite{madasgupta}.  The entanglement entropy at random-singlet critical points was already known~\cite{refaelmoore,laflorencie,dmcf-06,rm-07,refaelmoorereview} to show universal behavior similar to that at 1D conformal QCPs [Eq.~\eqref{criticalent}], with a modified prefactor of the logarithm (analogous to $c$) 
that was initially viewed as an effective central charge for random systems.

However, the results presented here indicate that this similarity does not extend to the full entanglement spectra, which 
are rather different.  We start by considering the disorder-averaged Renyi entropies
 \be
S_\alpha=\frac{1}{1-\alpha} \overline{\ln \tr{}{\rho_A^\alpha}}\, ,
\label{Sal}
\ee
where the bar denotes the average over quenched disorder.
These R\`enyi entropies $S_\alpha$
are quite simple in the random-singlet phase: they depend only on the mean number of singlets across the partition used to define 
the entanglement, just as does the entanglement entropy.  
The R\`enyi entropies already behave differently than in the conformal case.  
However, in disordered systems $S_\alpha$ is not the right quantity that determines the  entanglement spectrum via 
Laplace transform in $\alpha$ \cite{calabreselefevre}. 
To obtain the averaged moments of the distribution, one should instead consider 
the entropies corresponding to averaging the disorder {\it before} taking  the logarithm
\be
\widehat{S}_\alpha =\frac{1}{1-\alpha}\ln\overline{\tr{}{\rho_A^\alpha}}\, .
\label{Sahat}
\ee
This definition has also the advantage to maintain the relationship of the pure system between the Tsallis \cite{Ts} entropies 
${(\tr{}{\rho_A^\alpha}-1)}/{(1-\alpha)}$ and the R\`enyi entropies. 
These  
moments of the entanglement eigenvalue distribution reveal the full distribution of the number of singlets crossing a boundary and 
require an improved calculation.  
Both generalized entropies reduce to the Von Neumann one for $\alpha\to1$
\be
S_{VN}=\lim_{\alpha\to1}S_{\alpha}= \lim_{\alpha\to 1}\widehat{S}_{\alpha}\, .
\ee

The entropies $S_\alpha$ and $\widehat{S}_\alpha$ together with other properties 
are then studied for the random XX model and the validity of our results is discussed for general random-singlet  
ground-states. The manuscript is organized as follows. 
In Sec. \ref{S:rsp} we present the random-singlet picture and we derive the entropies $S_\alpha$ within 
strong disorder renormalization group. 
In Sec. \ref{S:pdf} we introduce the probability distribution of singlet formation and use it to derive the entropies 
$\widehat{S}_\alpha$.
Numerical tests of the predicted entropies and the discussion of their universality are described in Sec. \ref{S:num}.
Finally in Sec. \ref{S:con}, we report our main conclusions.

\section{Random-singlet picture of the Renyi entropies}
\label{S:rsp}

The ground state of a strongly disordered $s={1}/{2}$ Heisenberg chain or of the disordered XX chain
\be\label{eq:Hxx}
H=\frac{1}{4}\sum_l^LJ_l\Bigl(\sigma_l^x\sigma_{l+1}^x+\sigma_l^y\sigma_{l+1}^y\Bigr)\, ,
\ee
is described by the random-singlet phase (RSP) for essentially any probability distribution $P(J)$ of the coupling. When a system reaches this phase the ground state becomes almost factorized in singlets between spins at arbitrary large distances. The configuration of the singlets depends on the coupling constants \(J_l\), but several universal properties emerge in the average over disorder that are independent of the disorder distribution itself. The physical properties of a system in the RSP can be attained in an indirect way, \emph{i.e.} without referring (manifestly) to the particular Hamiltonian.
The real-space renormalization group approach (RSRG) is based on the picture that the strongest bond gives rise to a singlet, and the near-neighborhood spins can be described by means of an effective interaction from second-order perturbation theory.

Considering the XX Hamiltonian \eqref{eq:Hxx}, the Ma-Dasgupta rule \cite{fisher1} for the effective coupling constant after a decimation, \emph{i.e.} the formation of a singlet, is 
\be\label{eq:MDrule}
\Bigl(\cdots,J_l,J_M,J_r,\cdots\Bigr)_L\rightarrow \Bigl(\cdots,\frac{J_l J_r}{J_M},\cdots\Bigr)_{L-2}\, ,
\ee  
where \(J_M\) is the strongest bond of the chain of size \(L\) and \(J_l\) (\(J_r\)) is the near-neighborhood left (right) coupling constant. 
One of the most important consequences of \eqref{eq:MDrule} is that the distribution of the couplings 
after a sufficiently large number of decimations \(m\), with
\be
\beta_i^{(m)}=\ln\frac{J_M^{(m)}}{J_i^{(m)}},
\ee 
is substantially independent of the initial distribution:
\be\label{eq:attractor}
P(\beta)=\frac{1}{\Gamma^{(m)}}e^{-\frac{\beta}{\Gamma^{(m)}}}\, ,
\ee
where \(\Gamma\) is the RG flow parameter
\(
\Gamma^{(m)}=\ln\frac{J_M^{(0)}}{J_M^{(m)}}
\).
The distribution \eqref{eq:attractor} is the key to physical characteristics of the random-singlet phase. 
It is also the main ingredient for investigating the entanglement of spin blocks. 
In fact, for a spin block of length $\ell$ in a given RSP configuration with $n$
singlets linking the spins inside the subsystem with the spins outside (which we call in-out singlets)
the reduced density matrix is
\be
\rho_A^{RSP}\sim\bigotimes_{j=1}^n \begin{pmatrix}
\frac{1}{2}&0\\
0&\frac{1}{2}
\end{pmatrix}\bigotimes_{j=1}^{\frac{\ell-n}{2}} \begin{pmatrix}
0&0&0&0\\
0&\frac{1}{2}&-\frac{1}{2}&0\\
0&-\frac{1}{2}&\frac{1}{2}&0\\
0&0&0&0
\end{pmatrix}\, .
\ee
Thus, the entanglement of a subsystem of size $\ell$ depends only on the mean number $\overline n$ 
of in-out singlets.
In particular the entanglement entropy, as well as any Renyi entropy (\ref{Sal}),
is  proportional to the number of in-out singlets
\be\label{eq:SalphaRSP}
S_\alpha^{RSP}=\overline{n}\ln 2\, .
\ee
(This result has been also discussed in Refs. \onlinecite{refaelmoorereview} and \onlinecite{sc-10}.)
Ref.~\onlinecite{refaelmoore} shows that the averaged number of in-out singlets can be deduced directly from the flow equation for the distribution of couplings \(\beta_i\)
\begin{multline}
\frac{\mathrm d P(\beta)}{\mathrm d \Gamma}=
P(0)\int_0^\infty\mathrm d \beta_1\int_0^\infty\mathrm d \beta_2 \delta_{\beta-\beta_1-\beta_2}P(\beta_1)P(\beta_2)\\
+\frac{\partial P(\beta)}{\partial \beta} \, .
\end{multline} 
After some manipulation, this equation leads to \cite{refaelmoore}
\be
\overline{n}\simeq \frac{1}{3}\ln \ell\, ,
\ee
and so the entanglement entropy of a block of length $\ell$ is
\be\label{eq:SVNMoore}
S^{RSP}_{VN}(\ell)\simeq\frac{\ln 2}{3}\ln \ell\,,
\ee
with a weight-factor $\frac{\ln 2}{3}$ that calls to mind the behavior in the absence of disorder with an effective central charge $\ln 2$.

Consideration of the R\`enyi entropy rather than the standard entanglement entropy suggests that the similarity between 
the entanglement entropy with and without disorder is only superficial. Indeed in the RSP all R\`enyi entropies scale in the same way \eqref{eq:SalphaRSP}.  If we wish to define an effective central charge, we could use any conformal R\`enyi entropy \cite{cc-04}
\be
S_\alpha^{CFT}(\ell)=\frac{c}6 \left(1+\frac1\alpha\right) \ln \frac{\ell}{a}+c'_\alpha\,,
\label{Sacft}
\ee
as starting point, so that the effective central charge would have any value in the range $[(\ln 4)^{-1},(\ln 2)^{-1}]$ 
while $\alpha$ runs from $1$ to infinity. Also the central charge of the clean system $c=1$ belongs to this range, making 
questionable any attempt to generalize the Zamolodchikov ``c-theorem'' \cite{zamolodchikov}.
This picture from R\`enyi entropy is consistent with the previous counterexamples~\cite{santachiara,fidkowski} indicating that there 
is no version of the c-theorem for entanglement entropy that would describe the 
flow from clean to random systems~\cite{santachiara} or within random systems~\cite{fidkowski}.

The disorder-averaged R\`enyi entropies at random quantum critical points are universal and already indicate that the random-singlet phase's entanglement is quite different from the universal entanglement at 1D conformal QCPs.  However, since they depend on the same quantity (mean number $\overline{n}$ of in-out singlets) as the entanglement entropy, they do not probe new features of the random-singlet picture.  In the next section we consider additional quantities that are sensitive to new features and directly probe a memory effect in the RSRG flow, or ``repulsion between decimations'' in RG space, that was a key step in obtaining the correct value 
of $\overline{n}$.  Numerical tests of the predicted R\`enyi entropies are described in Section \ref{S:num}.

\section{Generalized entropy and the probability distribution of singlet formation}
\label{S:pdf}

The disorder-averaged Renyi entropy in the RSP only reflects the averaged number of the in-out singlets.   Thus it is not a natural measure of the full in-out singlet distribution \(P(n)\), or the probability distribution of the Renyi entropy. 
$P(n)$ can be examined by considering $\widehat{S}_\alpha$ in Eq. (\ref{Sahat}).
In fact, denoting with $g(t)$ the cumulant-generating function of the in-out singlet distribution $P(n)$
\be
g(t)=\ln\bigl<e^{n t}\bigr>\equiv\ln\sum_{n=0}^\infty P(n)e^{n t}\, ,
\label{gtdef}
\ee
it is straightforward that 
\be
\widehat{S}_\alpha^{RSP}=\frac{g(t(\alpha))}{1-\alpha}\,,
\label{Svsg}
\ee
where to keep the notation compact we defined 
\be
t= t(\alpha)\equiv (1-\alpha)\ln 2\,.
\label{tvsa}
\ee
Through all the paper $t$ will always denote this quantity, even when the $\alpha$ dependence is not specified.
$\widehat{S}_\alpha$ does depend on $\alpha$ in the RSP, unlike the R\`enyi entropy $S_\alpha$.
We require Eq. (\ref{gtdef}) to not blow up when $n\to\infty$, and so (assuming a reasonable $P(n)$) we need  
$t\leq0$ corresponding  to $\alpha\geq1$. We not not discuss a possible analytic continuation to $\alpha<1$ 
(that also in some clean systems can be complicated \cite{gt-10}).

From the real-space renormalization group (RSRG) point of view, singlets form at a constant rate with respect to a ``RG time'' \(\mu\), and this rate determines the logarithmic scaling of entanglement entropy.  En route to calculating this rate, Ref.~\onlinecite{refaelmoore} obtains the expression for the distribution of waiting times for a decimation across a bond since the last decimation:
\be\label{eq:distwaiting}
f(\mu)=\frac{1}{\sqrt{5}}\Bigl(e^{-\frac{3-\sqrt{5}}{2}\mu}-e^{-\frac{3+\sqrt{5}}{2}\mu}\Bigr)\, .
\ee
The above distribution has been deduced neglecting non-universal terms coming from the starting disorder distribution: Eq. \eqref{eq:distwaiting} is only asymptotically true. 
For example, we expect that the additive constant of the von Neumann entropy $S_{VN}$ should be disorder dependent.

During the RG time between two decimations several processes can happen. The most probable one is the formation of isolated singlets. Considering only this process leads to the renewal equation
\be\label{eq:Moorefond1}
\braket{e^{nt}}_\mu=\int_\mu^\infty\mathrm d \mu^\prime f(\mu^\prime) + e^t\int_0^\mu\mathrm d \mu^\prime f(\mu^\prime)\braket{e^{nt}}_{\mu-\mu^\prime}\, .
\ee
This equation can be solved by Laplace transformation.
Calling $\hat{f}(s)$ the Laplace transform of $f(\mu)$
\be
\hat{f}(s)=\frac{1}{\sqrt5}\left(\frac1{s+\frac{3-\sqrt5}2} -\frac1{s+\frac{3+\sqrt5}2}\right)\,,
\ee 
we have
\be
g_{(\mu)}(t)=\ln\Bigl[\mathscr{L}^{-1}\Bigl\{\frac{1}{s}\frac{1-\hat{f}(s)}{1-e^t \hat{f}(s)}\Bigr\}(\mu)\Bigr]\,,
\ee
and in particular $\displaystyle \overline{n}=\lim_{t\rightarrow 0^-}g'(t)$.

After simple algebra, we obtain 
\begin{multline}
\label{gt}
e^{ g_{(\mu)}(t)} = 
\left(\frac12 + \frac3 {2 \sqrt{5 + 4 e^t}} \right) e^{-\frac{3 -\sqrt{5 + 4 e^t}}2 \mu} \\
+ \left(\frac12 - \frac3{2 \sqrt{5 + 4 e^t}} \right) e^{-\frac{3 +\sqrt{5 + 4 e^t}} 2 \mu}\,,
\end{multline}
that via Eq. (\ref{Svsg}) gives $\widehat{S}_\alpha$ in the RSP. This is the main analytic result of this paper. 
It is useful to rewrite it in terms of the mean number of singlets as
\be
g(t)=  t A_t \overline{n}+t B_t\,,
\label{Mauform}
\ee   
where the multiplicative $t$ factor is introduced to write more compact formulas for $\widehat{S}_\alpha$ via Eq. (\ref{Svsg}). 
The two constants $A_t$ and $B_t$ are obtained 
by plugging \eqref{Mauform} into \eqref{gt}:
\be
\label{eq:Moore}
\left\{
\begin{aligned}
A_t&=3\frac{\sqrt{5+4e^t}-3}{2t }\,,\\
B_t&= \frac{1}{t}\ln\Bigl(\frac{1}{2}+\frac{3}{2\sqrt{5+4e^t}}\Bigr)+\frac{\sqrt{5+4e^t}-3}{6t }\, .
\end{aligned}\right.
\ee
Notice that in Eq. (\ref{Mauform}) all the dependence of $g_{(\mu)}(t)$ on $\mu$ is encoded in $\overline{n}$.
In this way, we also separated the universal $\ln \ell$ behavior (we remind $\overline{n}\propto \ln\ell$) given by 
$A_t$ from the constant one $B_t$.
We will come back to the discussion of the universal features of Eqs. (\ref{Mauform}) and (\ref{eq:Moore}) in the next section when 
comparing with the numerical results.

\section{Numerical Results}
\label{S:num}

In this section we present  numerical evidences confirming the critical scaling of the quantities 
calculated analytically by means of RSRG. We also present results for which we do not have yet any 
theoretical explanation, like the finite size scaling in the RSP.

The entropies $S_\alpha$ and $\widehat{S}_\alpha$ can be directly calculated for the disordered XX chain (\ref{eq:Hxx}), by 
generalizing the method of Laflorencie \cite{laflorencie}. 
In fact, for any realization of the disorder (i.e. any distribution of the bonds $J_l$), the XX model can be mapped into a
free-fermionic Hamiltonian by the Jordan-Wigner transformation $c^\dag_l=\prod_{j<l}\sigma_j^z \sigma^+_l$, that 
leaves the eigenvalues of the reduced density matrix of a single block unchanged, because the transformation is local inside the block. 
Defining the correlation matrix  $C_{l n}=\braket{c^\dag_l c_n}$,
the reduced density matrix of a spin block, that goes from the site $l_0+1$ to $l_0+\ell$, is the exponential of a free-fermion 
operator \cite{peschelfermion,ffo}
and it is completely characterized by the $\ell\times\ell$ correlation matrix $C$ in which indexes run from $l_0+1$ to $l_0+\ell$, 
that we call $C^{[l_0]}_\ell$. 
The entanglement entropy of the block in this configuration of the disorder is then given by 
\begin{multline}
S_{VN}^{[l_0]}(\{J_l\})=\\-\tr{}{C^{[l_0]}_\ell\ln C_\ell^{[l_0]}+(1-C^{[l_0]}_\ell)\ln(1- C_\ell^{[l_0]})}\,,
\end{multline}
while the Renyi entropy is
\be
S_\alpha^{[l_0]}(\{J_l\})=\frac{1}{1-\alpha}\tr{}{\ln\Bigl((C^{[l_0]}_\ell)^\alpha+(1-C^{[l_0]}_\ell)^\alpha\Bigr)}\, ,
\ee 
where we stressed the dependence on the disorder configuration $(\{J_l\})$ and on the first site of the block $l_0+1$. 
Indeed, on a single realization of the disorder, translational invariance is explicitly broken. 
Only after taking the disorder average translation symmetry can be restored. 
Having the R\`enyi entropies for a single realization, allows to obtain the asymptotic results for the disordered model, by 
averaging over a large enough number of configurations (generated randomly according to the specific rules for $\{J_l\}$).
$S_\alpha$ and $\widehat{S}_\alpha$ are obtained by averaging $S_\alpha$ or $e^{(1-\alpha)S_\alpha}$ respectively.
 
The method we presented is an {\it ab-initio} calculation of the R\`enyi entropies for disordered spin chains valid every time 
the model has a free-fermionic representation (as in XX or Ising chains). 
It is however numerically demanding. 
A more effective numerical technique exploits the RSP structure of the ground-state. 
Starting from a given disorder realization, we construct a singlet where the strong bond lies and we proceed to decimation 
according to the rule in Eq. (\ref{eq:MDrule}). We repeat this procedure until we spanned all the chain.
At this point we are left with a collection of singlets, and then, counting  number of singlets connecting the inside of the block with 
the outside, we have the configurational R\`enyi entropies from the relation $S_\alpha^{[l_0]}(\{J_l\})=n^{[l_0]}(\{J_l\}) \ln2$. 
As for the ab-initio calculation, $S_\alpha$ and $\widehat{S}_\alpha$ are obtained by averaging over the disorder. 
Note that $S_\alpha^{RSP}$ does not depend on $\alpha$ by definition, since for any configuration $S_\alpha=\overline{n} \ln2$.
Oppositely $\widehat{S}_\alpha$ depends on $\alpha$ because the average is taken over  $e^{(1-\alpha)S_\alpha}$ and indeed
some results for $\widehat{S}_\alpha$ have been already reported \cite{ss-10} by using this method.
For completeness, we give few  general features for an intuitive picture of the entanglement in the RSP.
After a decimation \eqref{eq:MDrule}, the renormalized bond is strongly suppressed, i.e. singlets repel.
The singlets that stay inside the block involve always an even number of spins, thus 
the parity of the block gives the parity of the number of in-out singlets. 
The spins belonging to the longest bonds crossing the two ends of the chain can be also thought as boundaries of two open chains. 
This suggests that in the RSP (as it is the case for clean systems \cite{cc-04}) the entanglement entropy 
of a block of $\ell$ spins in a periodic chain is equivalent to twice the entanglement entropy of ${\ell}/{2}$ spins in an open chain 
with the block starting from the boundary, i.e.  $S_\alpha^{\rm periodic}(\ell)\approx 2S_\alpha^{\rm open}(\ell/2)$. 
However,  this argument does not provide information about the additive constant (in 
clean models, the difference of the two constant terms gives the Affleck and Ludwig boundary entropy \cite{cc-04,al-91,otherbou}).

To avoid confusion between the two determinations of the entanglement, in the following we will always refer to 
the first method as {\it ab-initio} while to the second as RSP.
We stress that the RSP technique can be applied to any model with an RSP ground-state, as for example the disordered 
Heisenberg chains or spin-$1$ chains \cite{rm-07}, 
while the ab-initio one only to models having a free-femionic representation. 
However, the ab-initio method has the advantage to be exact by definition. 
Instead, by counting the number of singlets, we make the assumption that the ground-state has an RSP structure and that all 
the universal entanglement physics can be extracted from this. 
Although both assumptions sound reasonable, it is always worthwhile to perform in parallel the two numerical studies. 
In fact, the numerical counting of singlets is not the same as the analytic expressions derived in the previous
sections because, in order to provide analytic results, few further assumptions have been made (e.g. considering only the formation
of isolated singlets etc.). 
In case of disagreement between formulas and numerics, making the two computations in parallel  helps 
to understand if the error is in the approximations made to solve the equations or in the RSP assumption itself.

A possible generalization (that is currently under investigation \cite{f-prep}) is to 
understand if the RSP structure catches the entanglement of two disconnected blocks.
This can be achieved by calculating ab-initio R\`enyi entropies (indeed there are not known formulas for the 
entanglement entropy when the subsystem consists of more than one spin block, only some expressions  have been recently 
found for the first integer R\`enyi entropies  \cite{fc-10}) and comparing with the in-out singlets from both blocks.
It has been shown for conformal critical models \cite{2int,fc-10} that the entanglement of two blocks provides
much more information about the conformal structure than the single block one, and it is then worth investigating this issue  
also for the random case.

\subsection{Analysis of $S_\alpha$}

\begin{figure}[t]
\includegraphics[width=0.48\textwidth]{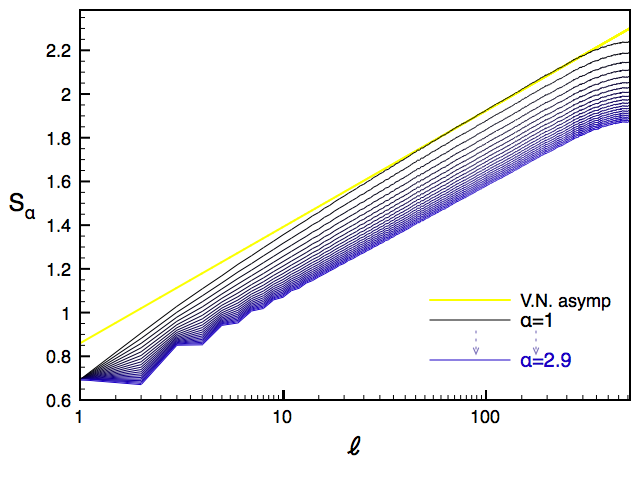}
\caption{Ab-initio Renyi entropies for a disordered XX chain of 1024 spins. 
The average is over 73000 realizations. 
The variation of the color shows results from $\alpha=1$ (upper line) to $\alpha=2.9$ (bottom line). 
The yellow line is the asymptotic Von Neumann entropy ($\alpha=1$) 
obtained by Laflorencie \cite{laflorencie}.
}
\label{fig:S1024}
\end{figure}

We computed ab-initio the averaged R\`enyi entropies $S_\alpha$ for many different system sizes. 
In Fig. \ref{fig:S1024}, we report the result for a chain of $L=1024$  spins for the disorder average over a sample of $73000$ 
realizations. 
For $1\ll\ell\ll L$, the various curves are parallel, with the slope predicted by Eq. (\ref{eq:SVNMoore}), i.e. 
the leading term of $S_\alpha$ is $\alpha$ independent.
The non-universal additive $O(1)$ term clearly depends on $\alpha$, as in the clean case.
On top of a smooth behavior, we can see oscillating contributions,  evident for small $\ell$ and large $\alpha$. 
Their presence does not come unexpected: also in clean chains \cite{ccen-10,cc-10,otherosc}, 
there are oscillating terms that  (in zero magnetic field) are  parity dependent, i.e. they are of the form $(-1)^\ell$. 
However,  for random systems the oscillations have a different form and they decay 
rather quickly with $\ell$ (as opposite to $\widehat{S}_\alpha$ as we shall see).
We do not have a proper theory for their origin, nor a phenomenological description, but
their understanding is  beyond the goals of this paper since they do not influence the determination of the asymptotic bahavior. 
When $\ell$ approaches the chain length $L$, sizable finite-size corrections are visible.
Next subsection will be devoted to their accurate study, while here we continue with the asymptotic analysis of $S_\alpha$.

\begin{figure}[t]
\includegraphics[width=0.48\textwidth]{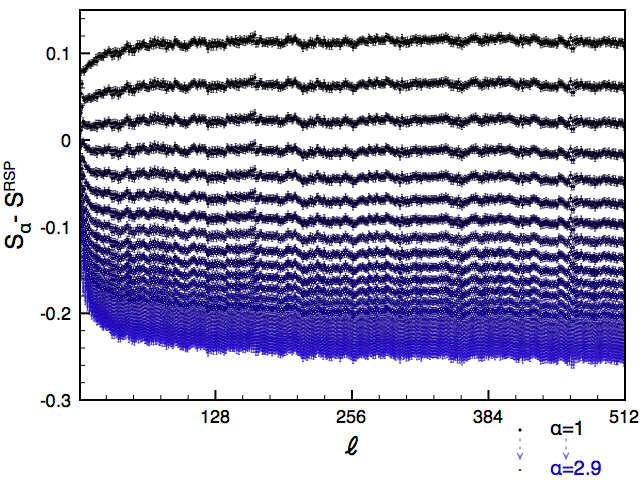}
\caption{Ab-initio R\`enyi entropies for a disordered XX chain of 1024 spins minus the RSP value.
The averages are over the same sample of 73000 realizations.
}\label{fig:S1024subSP}
\end{figure}

We compare the data in Fig. \ref{fig:S1024} from the ab-initio calculation, 
with the numerical results obtained using the RSP approach on the same random sample of 73000 realizations of $J_l$.
According to Eq. \eqref{eq:SalphaRSP}, the RSP R\`enyi entropies do not depend on $\alpha$ by definition.
For this reason,  in Fig. \ref{fig:S1024subSP} we report the difference between the RSP  R\`enyi entropies and the 
ab-initio ones presented in Fig. \ref{fig:S1024}.
After a transient behavior for small $\ell$, all the curves with varying $\alpha$ approach a constant, indicating not only that   
the universal leading logarithmic term in $S_\alpha$ is correctly described by RSP, but also the
finite size corrections are.
In the range of  $\alpha$ considered in the figure, we find that the additive constant is well described by 
\be
S_\alpha\approx S^{RSP}_\alpha+\frac{a}{\alpha}+b+o(1)\, ,
\ee 
where the disorder-dependent constants $a$ and $b$ in the case of random disorder take the values
$a\approx 0.61$ and  $b\approx -0.47$.

\subsection{Finite-size effects}

Having established the correctness of the asymptotic RSRG results for $S_\alpha$ in the region $1\ll\ell \ll L$, we can 
consider the finite-size effects. 
One of the most remarkable result of conformal invariance is that the finite-size 
scaling is obtained with the replacement
\be
\ell\rightarrow \frac{L}{\pi}\sin\Bigl(\frac{\pi\ell}{L}\Bigr)\, .
\ee
in the thermodynamic limit result. The rhs of the above equation is known as chord length.
However, when conformal invariance is broken, the chord length does not give the finite-size scaling.
In fact, using the results reported above, it is easy to show that this is the case, 
as it was already shown for some random Ising systems \cite{il-08}.

\begin{figure}
\includegraphics[width=0.48\textwidth]{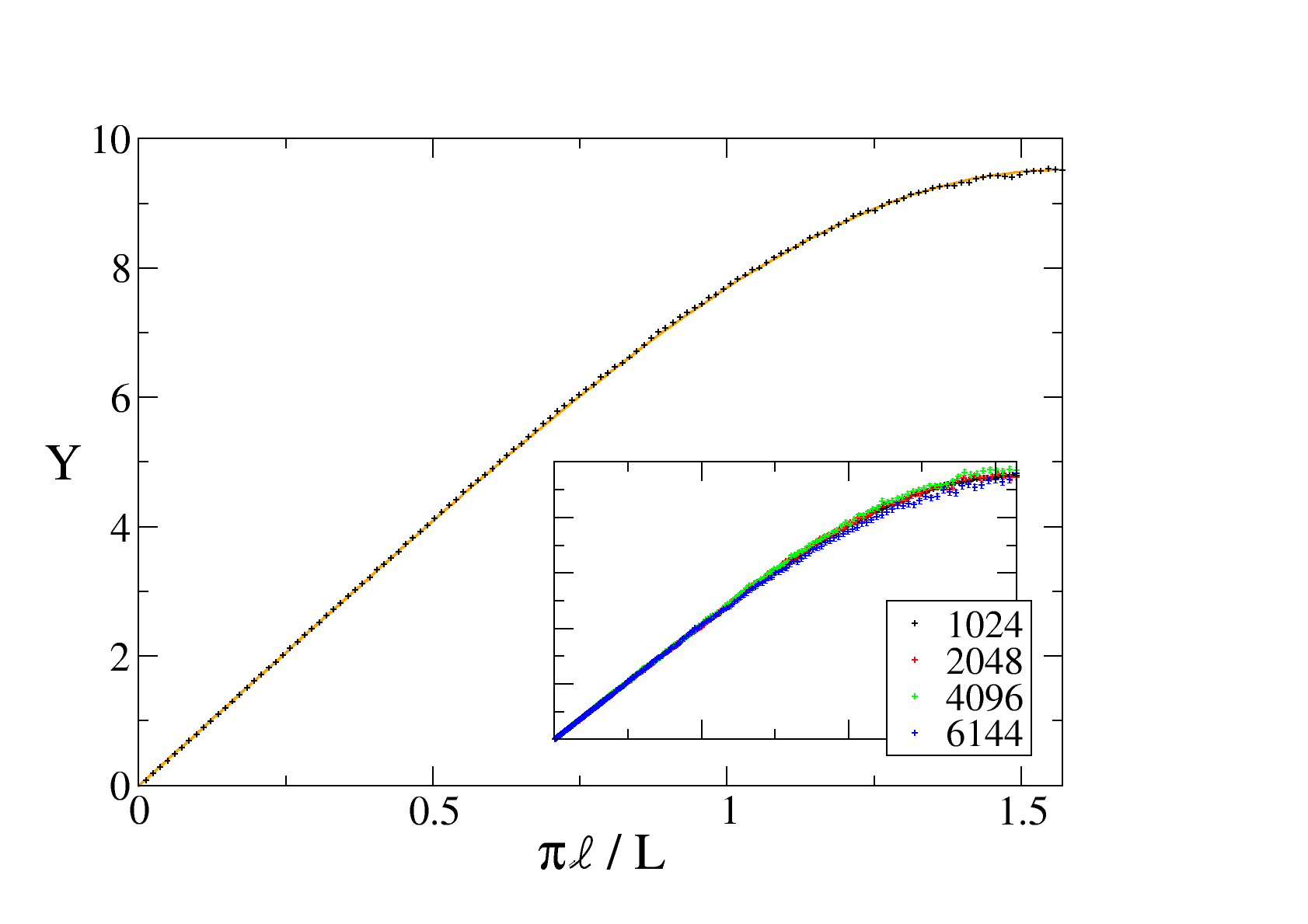}
\caption{The finite-size scaling function for the entanglement entropy $Y(x)$ in Eq. (\ref{YY}).
Main: RSP data averaged over 1440000 disorder realizations for $L=1024$. 
The continuous (red) curve is the proposed phenomenological formula (\ref{Yp}) describing perfectly the data points.
Inset: The same plot for different values of $L$, showing the collapse on a single scaling function.
}\label{fig:chordlength}
\end{figure}

Even if conformal invariance is broken, scale invariance still holds. 
Thus the finite size scaling can always be taken into account by the substitution
\be
\ell\rightarrow \frac{L}\pi\, Y\left(\frac{\pi \ell}L\right)\,.
\label{YY}
\ee
The great predictive power of conformal symmetry is that independently of the observable (but built with primary operators) the scaling 
function is always $Y(x)=\sin(x)$, while in general scale-invariant theories the function $Y(x)$ {\it does depend on the observable}.
Some results on the finite-size scaling of entanglement in 1D critical non-conformal systems have  been already 
reported \cite{il-08,ps-05,ar-10,cv-10}.
The function $Y(x)$ for $S_\alpha$ must however satisfy simple symmetry constraints. 
First, $S_\alpha$ is symmetric for $\ell\to L-\ell$, thus $Y(x)= Y(\pi-x)$. 
Second, periodic boundary conditions require $S_\alpha$ to be a periodic function of $\ell$ of period $L$, and so 
$Y(x)=Y(\pi+x)$. Thus we can expand $Y(x)$ in Fourier modes as
\be
Y(x)=\Bigl[1+\sum_{j=1}^\infty {k_j} \Bigr]\sin x-\sum_{k=1}^\infty\frac{k_j}{2j+1}\sin ((2j+1)x)\, ,
\ee
where we also imposed $Y(x\ll1)\sim x$ to reproduce the correct thermodynamic limit.
The chord length has only the first mode and so corresponds to $k_j=0$ for any $j$.
This expansion in terms of Fourier modes is particularly useful, because we expect 
that the contribution of the first few modes will be enough to have a reasonable approximation of the scaling function $Y(x)$.
Indeed, Fig. \ref{fig:chordlength} shows that only the first term $k_1$ is  enough to describe accurately the observed behavior
for the RSP entanglement entropy
\begin{eqnarray}
Y(x)&\simeq& (1+k_1)\sin x-\frac{k_1}{3}\sin 3x
\nonumber \\&=&
\sin x \Bigl[1+\frac{4}{3}k_1\sin^2 x\Bigr]\,, 
\label{Yp}
\end{eqnarray}     
with $k_1\approx 0.115$.
The obtained scaling function in presence of disorder is greater than the chord length.

Fig. \ref{fig:S1024subSP} shows that the finite-size scaling in the ab-initio calculation are equivalent to the RSP ones (else for 
$\ell\sim L$ the various curves should bend). This means that the finite-size scaling of all $S_\alpha$ in the spin chain is described by 
Eq. (\ref{Yp}), as we also checked directly.

\subsection{Probability distribution of the R\`enyi entropy}

\begin{figure}
\includegraphics[width=0.48\textwidth]{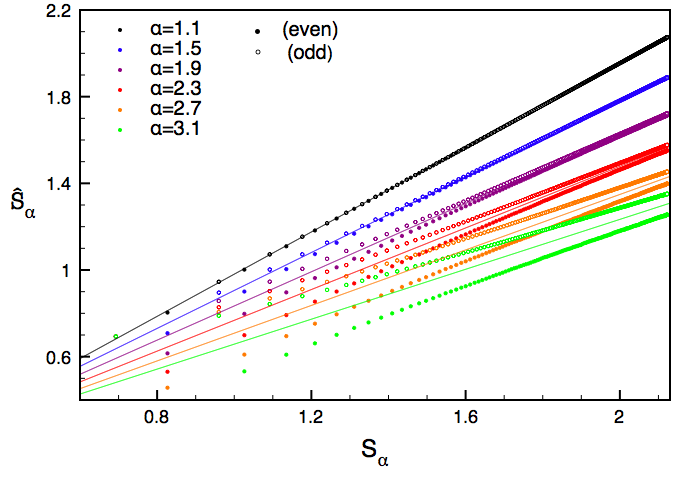}
\includegraphics[width=0.48\textwidth]{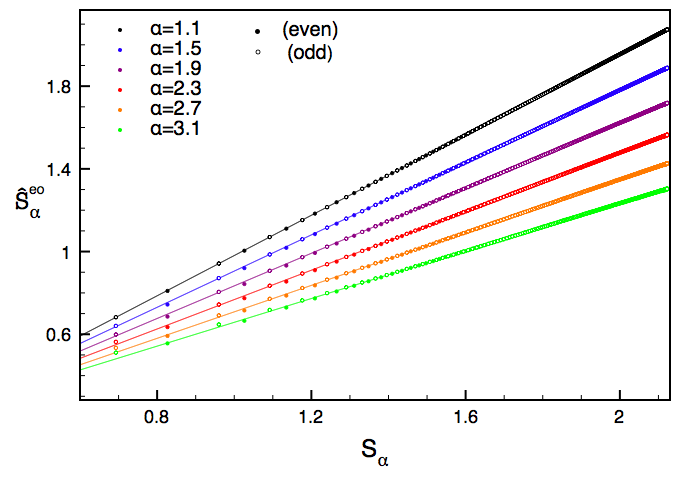}
\caption{Top: RSP results for $\widehat{S}_\alpha$ as a function of $S_{\alpha}$ for a chain of $1024$ spins and 1440000 
disorder realizations. 
Bottom: Even-odd average of $\widehat{S}_\alpha$ eliminating leading corrections to the scaling. 
In both panels, the continuous lines are the analytic RSRG result for $A_t$.
}\label{fig:SstarS}
\end{figure}

The disorder averaged R\`enyi entropy $S_\alpha$ gives only access to the averaged number of the in-out 
singlets, while  $\widehat{S}_\alpha$ gives access to the full  in-out singlets distribution $P(n)$, i.e. the probability distribution of 
the R\`enyi entropy and so to the full entanglement spectrum. 
Indeed  $\widehat{S}_\alpha$ is related to the the cumulant generating function 
$g(t) $ of the in-out singlets distribution by Eq. (\ref{Svsg}).

We first consider the RSP data, because they allow to explore larger system sizes. Only after having established the  
asymptotic behavior we will consider ab-initio data and show consistency with the proposed scaling.

We  observed that the R\`enyi entropies $S_\alpha^{\rm RSP}$ do not have subleading corrections depending 
on the parity of the block, making the asymptotic analysis quite straightforward.
Oppositely, the data for $\widehat{S}_{\alpha}^{RSP}$ (see Fig. \ref{fig:SstarS}) show that they depend on the block parity 
in a way similar to clean systems \cite{ccen-10}. 
To analyze the numerical data we conjecture the following asymptotic behavior
\begin{multline}
\widehat{S}_\alpha^{RSP}(\ell)\approx \\ A_t S_{\alpha}^{RSP}(\ell)+ B_t \ln 2-(-1)^{\ell} f_t\bigl(S_{\alpha}^{RSP}(\ell)\bigr)\ln 2\, ,
\label{Swithcorr}
\end{multline}
where $t$ is defined in  Eq. (\ref{tvsa}). 
$A_t$ and $B_t$ are the two functions introduced in Eq. (\ref{Mauform}), while $f_t$ takes into account the corrections
to the scaling and goes to $0$ for $\ell\to\infty$.
The form of the corrections is inspired by the results in clean systems, while the leading term is the solution asymptotic $g(t)$
in Eqs. (\ref{Mauform}) and (\ref{eq:Moore}).
In the top of Fig. \ref{fig:SstarS} we also report the RSRG value for $A_t$ that seems to be in qualitative agreement with 
the numerical data. A full quantitative description requires the elimination of the corrections to the scaling.

In order to provide an unbiased description of the asymptotic behavior of $\widehat{S}_\alpha$, 
we define the functions  $s_\alpha^{\rm even}(\ell)$ and  $s_\alpha^{\rm odd}(\ell)$ 
from the interpolation relative to even and odd blocks respectively. 
We can isolate the leading behavior of $\widehat{S}_\alpha^{RSP}$  by considering the average over the two interpolating functions, i.e.
\be
\widehat{S}_\alpha^{eo}(\ell)\equiv
\frac{s_\alpha^{even}(\ell)+s_{\alpha}^{odd}(\ell)}{2}\, .
\ee
This definition eliminates  the leading corrections to the scaling.
In fact, in the lower panel of Fig. \ref{fig:SstarS} we have a linear relation between $\widehat{S}_\alpha^{eo}$ and
$S_\alpha^{RSP}$  for all reported values of $\alpha$ 
(while the non-averaged data in the top panel are linear only for $\alpha$ close to $1$).

\begin{figure}[b]
\includegraphics[width=0.48\textwidth]{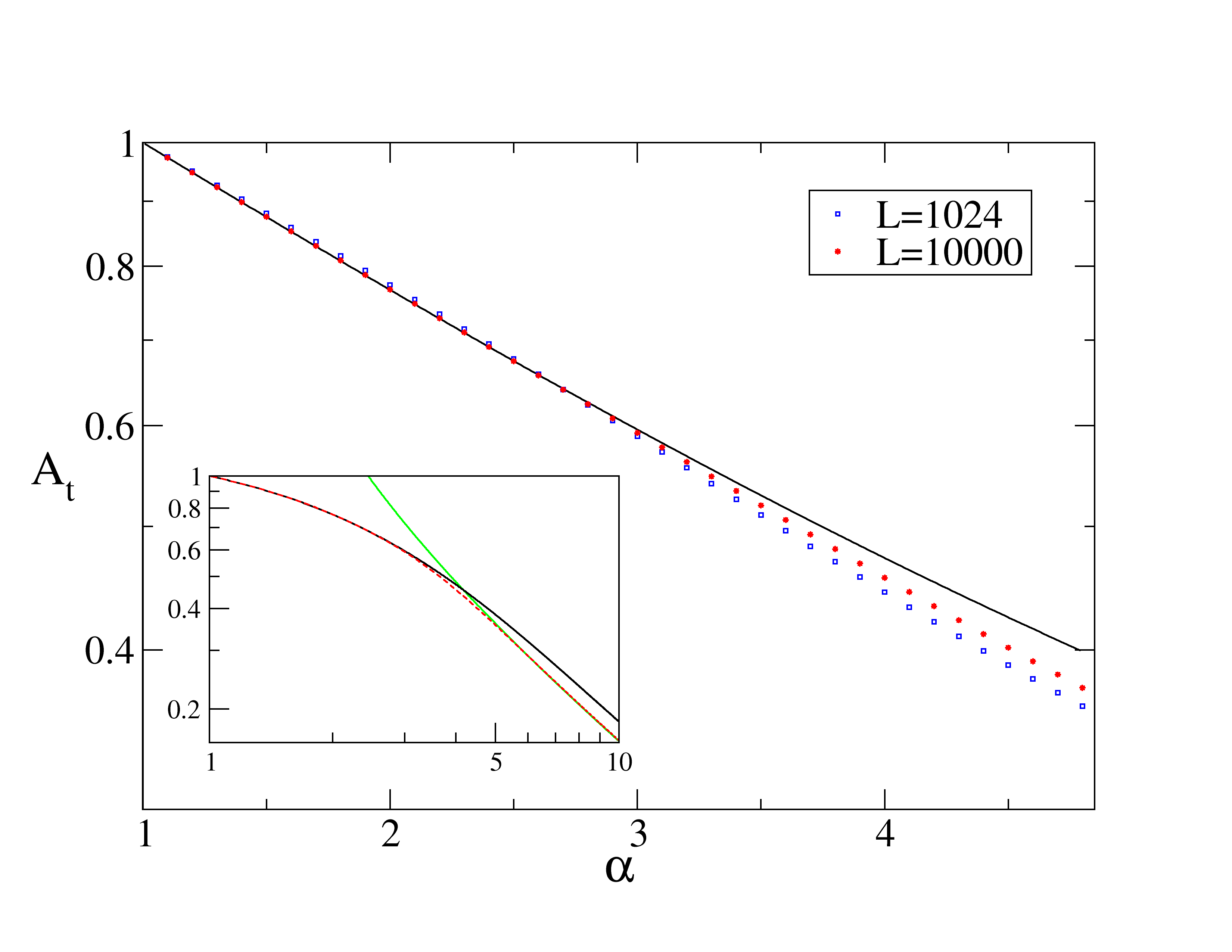}
\caption{The universal constant $A_t$ obtained from RSP data for $L=1024$ (1440000 disorder realizations) and $L=10000$
(320000 realizations).
Main plot: For $\alpha\leq3.5$ finite-size effects are negligible and the RSRG prediction (continuous line) describes the data.
Inset:  Crossover  to the non-universal Poissonian behavior (green continuous line) for larger $\alpha$.
}\label{fig:At}
\end{figure}

From this linear dependence we can extract the functions $A_t$ and $B_t$ using the RSRG relation
\be
\widehat{S}_\alpha^{eo}\simeq A_t S_{\alpha}^{RSP}(\ell)+\ln 2 B_t\,.
\ee
The resulting values for the universal coefficient $A_{t(\alpha)}$ for $\alpha\leq 10$ and for $L=1024$ and $L=10000$
are reported in Fig. \ref{fig:At}.
For small $\alpha$ ($\leq 3.5$) there are negligible fiinite-size corrections and the data perfectly agree with the RSRG 
result in Eq. (\ref{eq:Moore}), showing the predictive power of the RSRG to determine $A_t$.
For larger $\alpha$, finite-size corrections are important and indeed the data differ from the analytical prediction, but the 
larger system sizes are closer. We believe that in the thermodynamic limit the RSRG $A_t$ describes the correct behavior for 
any $\alpha$. The reason of these finite-size effects is also easily understood: the asymptotic formula is valid for 
$\widehat{S}_\alpha$ large, while in this region of $\alpha$ we have $\widehat{S}_\alpha\sim 1$.
Even if not asymptotic, the large $\alpha$ results show an interesting behavior: independently of $L$, they follow
a $-1/t$ behavior (see inset in Fig. \ref{fig:At}), typical of a Poissonian distribution of singlets.
The reason of this Poissonian behavior can be traced back to the fact that for $t\to-\infty$ we are giving a large weight 
to short-range singlets that are produced almost independently. 
Little weight is instead given to long-range singlets responsible of the universal physics and so for these 
values of $\alpha$ and $L$ we are probing the UV physics.
According to this interpretation, a crossover from the universal behavior  of Eq. (\ref{eq:Moore}) to a UV Poissonian behavior
always takes place for $\alpha\sim\ln L$, in  agreement with Fig. \ref{fig:At}.

\begin{figure}[t]
\includegraphics[width=0.48\textwidth]{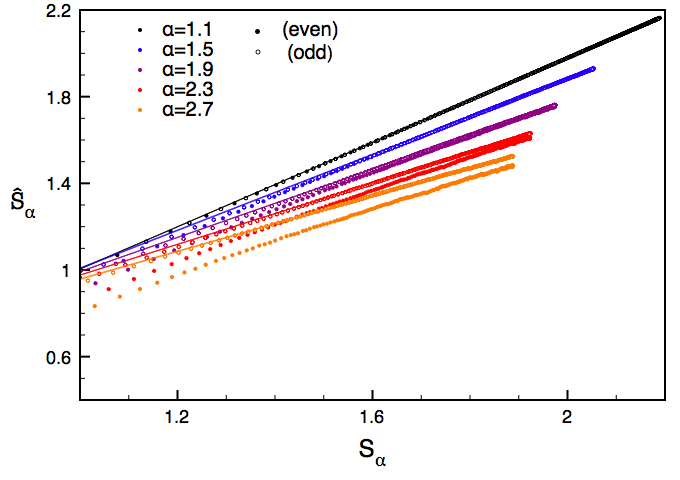}
\caption{Ab-initio $\widehat{S}_\alpha$ as a function of $S_{\alpha}$ for a spin-chain of $1024$ spins  and 73000 disorder realizations. 
The continuous lines represent the RSRG prediction for the slope.
The additive terms are different from those in Fig. \ref{fig:SstarS}.
}\label{fig:SstarSchain}
\end{figure}

We can now move to the ab-initio calculation to check the validity of the RSP scenario for $\widehat{S}_\alpha$.
As before,  we focus on the relation between $\widehat{S}_\alpha$ and $S_\alpha$ and
in particular on the universal slope of the linear relation between them.
The results are reported in Fig. \ref{fig:SstarSchain}. 
Asymptotically, the slopes of these curves tend to the RSRG prediction for  $A_t$ shown as continuous lines in the figure. 
Also the finite-size scaling scaling is well described by Eq. (\ref{Yp}), as evident from the 
fact that the linear relation between $\widehat{S}_\alpha$ and $S_\alpha$ is correct even for large values of $\ell$ (i.e. of $S_\alpha$)
in the various plots.
However, as clear by a visual comparison between Figs. \ref{fig:SstarSchain} and  \ref{fig:SstarS} (top), 
the constant term in this relation is different (and both different from the analytic $B_t$ in Eq. (\ref{eq:Moore})). 
The degree of universality of this term is discussed in next subsection.

\begin{figure}[t]
\includegraphics[width=0.48\textwidth]{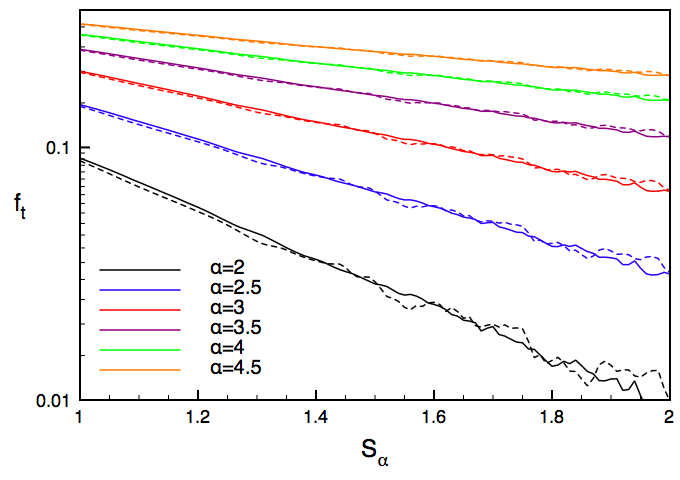}
\caption{Scaling functions for the correction to the scaling $f_t(S)$ in Eq. (\ref{Swithcorr}) obtained as difference of 
$s_\alpha^{\rm odd}(\ell)$ and  $s_\alpha^{\rm even}(\ell)$. Full and dashed lines correspond to uniform and exponential distributions
of disorder respectively.}
\label{fig:corr}
\end{figure}

Having established the correct asymptotic behavior we can consider the oscillating corrections to the scaling defined in 
Eq. (\ref{Swithcorr}).  
The numerical estimate of $f_t(S)$ can be obtained as
\be
f_t(S)\simeq  \frac{s_\alpha^{\rm odd}(\ell)-s_\alpha^{\rm even}(\ell)}2+\dots\,,
\ee
where the dots denotes subsubleading terms (we recall $s_\alpha^{odd/even}$ are interpolations and so defined for any $\ell$).
The data obtained in this way are reported in Fig. \ref{fig:corr}. 
The linear behavior in log-scale shows that for $\alpha\leq5$ 
(for larger $\alpha$ further sub-leading corrections must be considered \cite{ccen-10}) 
$f_t(S_\alpha)$ decays exponentially
\be
f_{t}(x)=F_te^{-\nu_t x}\, ,
\ee
i.e. a power-law correction in $\ell$. 
$\nu_{t(\alpha)}$ is a new universal critical exponent governing the corrections to the scaling of $\widehat{S}_\alpha$, 
analogous to the one introduced in clean systems \cite{ccen-10,cc-10}.
We can see that $\nu_{t(\alpha)}$ decreases with increasing $\alpha$, but
a precise numerical estimate is difficult. 
For clean systems it has been shown\cite{ccen-10,cc-10} that $\nu_{t(\alpha)}=2K/\alpha$ with $K$ an $\alpha$-independent 
exponent equal to the scaling dimension of a relevant operator. 
We can rule out this form for the random spin-chain, but
the accuracy of our results does not allow to establish numerically an exact formula for the $\alpha$ dependence of the exponent.
We also mention that the corrections to the scaling are of the same form also in ab-initio calculations, 
as qualitatively clear from Fig. \ref{fig:SstarSchain} and quantitatively checked but not reported here. 
This shows  the correctness of the RSP description and also that the real spin-chain does not introduce new
leading corrections to the scaling in addition to the RSP ones.

\begin{figure}[b]
\includegraphics[width=0.48\textwidth]{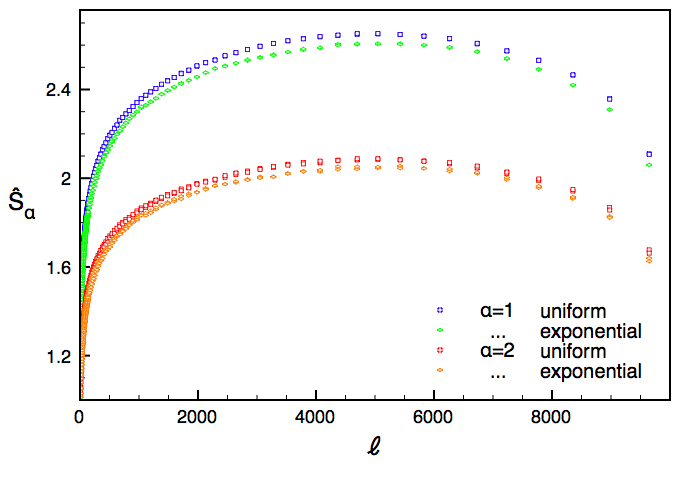}
\caption{$\widehat{S}_\alpha$ for two disorder distributions. 
The RSP data are for chains of 10000 spins and averaged over 320000 configurations. 
}\label{fig:addcnt}
\end{figure}

\subsection{Universality}

All the results presented until now, both ab-initio and RSP have been obtained for random distributions of the coupling constant 
$J$ in the interval $[0,1]$. However, the universal prediction of RSRG must be independent of the distributions of $J$ (as long 
as new symmetries are not introduced).
We check this universality by studying the RSP chain with $L=10000$ spins with coupling distributed both uniformly $J\in[0,1]$ 
and exponentially $P(J)\sim e^{-J}$.
In Fig. \ref{fig:addcnt} we report the numerical RSP results of $\widehat{S}_\alpha$ for $\alpha=1,2$ for the two distributions
of the disorder.
As expected, the two distributions lead to slightly different results: only the leading logarithmic term in $\ell$ is universal, 
while the additive constant term is not.

\begin{figure}[t]
\includegraphics[width=0.48\textwidth]{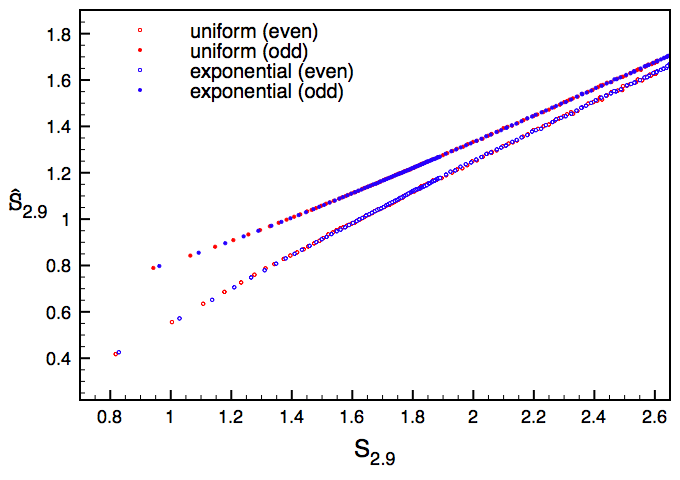}
\caption{$\widehat{S}_\alpha$ as function of $S_\alpha$ for $\alpha=2.9$ and for two disorder distributions (RSP data with $L=10000$
and 320000 configurations). The scaling function is disorder independent.
}\label{fig:SastvsSVN}
\end{figure}

To check the universality of the leading term, Fig. \ref{fig:SastvsSVN}  reports  
$\widehat{S}_\alpha$ as function $S_\alpha$ for $\alpha=2.9$ (other values of $\alpha$ leads to equivalent plots) 
for the two distributions. 
The two curves perfectly coincide, despite when they are plotted as function of $\ell$ they are different.
This means that all the non-universal behavior of the additive constants is washed out and we are left with a universal function.
At first this result can seem surprising, 
but it is easy to realize that, in this kind of plots, the dependence on the non-universal cut-off or 
lattice scale $a$ disappear and the leftover difference of non-universal additive constants is universal. 
For example, for the conformal entropies (\ref{Sacft}) we have the {\it universal} relation
\be
S_\alpha^{CFT}=\frac{S_{VN}^{CFT}}2\left(1+\frac1\alpha\right)+c'_\alpha-\frac{c'_1}2\left(1+\frac1\alpha\right)\,,
\label{univ}
\ee
where evidently all the $a$ dependence disappeared. 
To our knowledge, this property has not been explored at all in clean systems, but e.g. one can easily check that in the exact results 
for the critical XY model\cite{jk-04}, the dependence on the irrelevant parameter $\gamma$ disappears in Eq. (\ref{univ}).

\begin{figure}[t]
\includegraphics[width=0.47\textwidth]{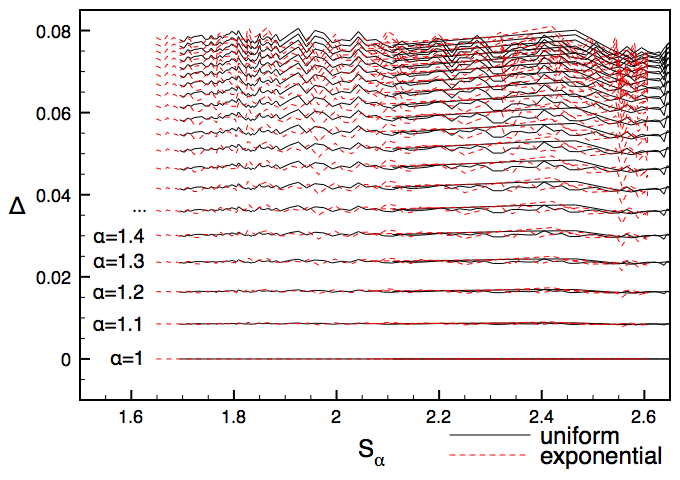}
\caption{The quantity  $\Delta$ defined in Eq. (\ref{Delta}) vs  $S_\alpha$ for uniform and exponential distributions of
disorder. With varying $\alpha$ the two differences are the same, showing the universality of the coefficient $B_{t(\alpha)}$.
}\label{fig:DeltaMoore}
\end{figure}

Having established that both $A_t$ and $B_t$ are universal, 
we reconsider our results for the disordered systems. 
We already discussed for the uniform distribution (see Fig. \ref{fig:At}) how the numerical value of $A_t$ 
agrees with the analytical RSRG prediction. The independence of  $A_t$ on the disorder distribution confirms its
universality. 
In Fig. \ref{fig:DeltaMoore}, 
we plot the quantity 
\be
\Delta=\widehat{S}^{eo}_\alpha-\frac{g_\mu (t(\alpha))}{1-\alpha}\,,
\label{Delta}
\ee
where $g_{(\mu)}(t)$ is the function in Eq. (\ref{gt}) and $\mu$ is fixed by $S_\alpha$ via
$\mu= {\frac{3}{\ln 2}S_\alpha+\frac{1}{3}}$. This quantity has been built in such a way to cancel the leading behavior 
$A_t$ so to leave only $B_t$. Albeit little noisy, Fig. \ref{fig:DeltaMoore} shows clearly the disorder independence of $B_t$.

Disappointingly, as shown for uniform disorder, 
we found that the RSP and ab-initio calculation for $\widehat{S}_\alpha$ provide different values for the 
constant $B_{t(\alpha)}$ that are both different from the RSRG expression in Eq. (\ref{eq:Moore}).
On one hand, this is showing that the RSP description is unable to catch this feature of the spin-chain because numerical RSP
and ab-initio data disagree. 
On the other hand, this is also showing that while carrying out the analytic results for $g(t)$, some of the assumptions made 
influence significantly this quantity.      
There are two possible explanations to motivate the last discrepancy.
One is that the distribution $f(\mu)$ in Eq. \eqref{eq:distwaiting} contains some additional (subleading) terms not considered here.
In fact, as already discussed, Eq. \eqref{eq:distwaiting}  has been deduced neglecting terms coming from the starting disorder 
distribution and it is only asymptotically true. 
The other possibility is instead that the discarded terms in the renewal equation \eqref{eq:Moorefond1} contribute to $B_t$. 
Several pieces of information have been indeed ignored there: 
memory beyond first order,  multiple decimations, the flow of the distribution to the critical point, etc. 
We found it rather improbable that $f(\mu)$ should be modified. It is difficult to imagine how to modify it keeping 
all the other correct results (i.e. the entanglement entropy, $A_t$ etc.). 
On the other hand, solving the renewal equation in the presence of the discarded effects is very hard (maybe impossible). 
Thus, to convince ourself that these processes can be responsible of a changing in $B_t$, we tried to add some 
oversimplified processes (but physically motivated) to the renewal equation:
we found that all these processes change $B_t$, but leave
$A_t$ unchanged, showing that this is the most probable explanation of the discrepancy.
However, from the ab-initio results, we know that the real spin-chain introduces further corrections to this term
and so we do not find reasonable to embark in a difficult calculation, that in any case will not provide the correct answer for the 
spin-chain. 

To conclude the universality section, it is worth to mention that the oscillating corrections to the scaling (the function $f_t$ in  
Eq. (\ref{Swithcorr})) also do not depend on the disorder distribution as shown in Fig. \ref{fig:corr}, confirming their universality.

\section{Conclusions}
\label{S:con}

We provided an analytical and numerical description of the  R\`enyi entropies 
$S_\alpha$ and $\widehat{S}_\alpha$ in a random singlet phase. 
For $S_\alpha$ the leading logarithmic behavior is $\alpha$-independent and only the subleading constant term 
depends on $\alpha$:
\be
S_{\alpha}=\frac{1}{1-\alpha}\overline{\ln\tr{}{\rho^\alpha}}\simeq \frac{\ln 2}{3} \ln \ell +d'_\alpha
\ee
The leading universal term has been determined analytically, while the non-universal correction $d'_\alpha$ only numerically.
Oppositely, the leading universal term of  $\widehat{S}_\alpha$ has a non-trivial $\alpha$-dependence.
Its scaling behavior can be written in a completely {\it universal} form as
\be
\widehat{S}_\alpha=\frac{1}{1-\alpha}\ln\overline{\tr{}{\rho^\alpha}}\simeq A_{(1-\alpha)\ln 2}S_\alpha+B_{(1-\alpha)\ln 2} ,
\ee
Indeed, we pointed out that the functions  $A_t$ and $b_t$ connecting linearly $\widehat{S}_\alpha$ and $S_\alpha$ are both 
independent of the cut-off length introduced by the chain, and so universal.
The analytic result based on the solution of real-space renormalization group equations agrees perfectly with the numerical data 
as shown in Fig. \ref{fig:At}, giving a full characterization of the asymptotic behavior.
Instead a first-order RG prediction for the subleading term $B_t$ disagrees with the numerical data. 
We showed evidences that this disagreement should be related to the approximations done in the RG equations. 
Only an improved, but much more difficult (and maybe impossible) calculation can provide the exact result for $B_t$.

We also studied the finite-size scaling:
for finite chains the above relations still hold if the subsystem length $\ell$ is replaced by a {\it modified chord length} that is
phenomenologically well approximated by Eq. (\ref{Yp}). We do not have a theoretical explanation for this finite-size scaling form.

Assuming that the random-singlet description is equally valid for the random Heisenberg model, as is plausible and often assumed 
but not yet proved or firmly confirmed numerically, then we are in the surprising situation of knowing the entanglement spectrum 
exactly for the random Heisenberg model but only approximately for the corresponding pure model 
(apart from some exact results for small $\ell$ \cite{xxzv}).


\end{document}